\documentclass[8pt]{revtex4}
\pdfoutput=1
\usepackage{amssymb}
\usepackage{latexsym}
\usepackage{epsfig}
\begin{document}                             

\title{ Cosmology from quantum potential in a system of oscillating branes }

\author{ Alireza Sepehri $^{1,2}$\footnote{alireza.sepehri@uk.ac.ir}, }
\address{ $^1$ Faculty of Physics, Shahid Bahonar University, P.O. Box 76175, Kerman, Iran.\\$^{2}$
Research Institute for Astronomy and Astrophysics of Maragha
(RIAAM), Maragha, Iran.}

\begin{abstract}
Recently, some authors proposed a new mechanism which gets rid of
the big-bang singularity and shows that the age of the universe is
infinite. In this paper, we will confirm their results and predict
that the universe may expand and contract many times in a system
of oscillating branes. In this model, first, N fundamental strings
transit to N M0-anti-M0-branes. Then, M0-branes join to each other
and build a M8-anti-M8 system.  This system is unstable, broken
and two anti-M4-branes, a compactified M4-brane, an M3-brane in
additional to one M0-brane are produced. The M3-brane
 wraps around the compactified M4-brane and both of them oscillate between two anti-M4-branes. Our universe is located on
 the M3-brane and interacts with other branes by exchanging the M0-brane and some scalars in transverse directions.
By wrapping of M3-brane, the contraction epoch of universe starts
and some higher order of derivatives of scalar fields in the
relevant action of branes  are produced which are the responsible
of generating the generalized uncertainty principle or GUP. By
oscillating the compactified M4-M3-brane and approaching to one of
anti-M4-branes, one end of M3-brane glues to the anti-M4-brane and
other end remains sticking and wrapping around M4-brane. Then, by
getting away of the M4-M3 system, M4 rolls, wrapped M3 opens and
expansion epoch of universe begins. By closing the M4 to anti-M4,
the square mass of some scalars become negative and they make a
transition to a tachyonic phase.  To remove these states, M4
rebounds, rolls and M3 wraps around it again. At this stage,
expansion branch ends and universe enters to a contraction epoch
again. This process is repeated many times and universe expands
and contracts as due to oscillating of branes. We obtain the scale
factor of universe in this system and find that it's values only
at $t\rightarrow -\infty$ shrinks to zero. Thus, in our method,
the big bang is replaced by the fundamental string and the age of
universe is predicted to be infinite. Also, when tachyonic states
disappear at the beginning of expansion branch, some extra energy
is produced and leads to an increase in the velocity of opening of
M3. In these conditions, our universe which is located on this
brane, expands very fast and experiences an inflation epoch.
Finally, by reducing the fields in eleven dimensional M-theory to
the fields in four dimensional universe, we show that our theory
matches with quantum field theory prescriptions.

\end{abstract}

 \maketitle
\section{Introduction}

Removing different singularites from cosmological models is one of
important problems in physics that many scientists are trying to
solve it. Recently, it has been shown that replacing classical
trajectories or geodesics by their quantal (Bohmian) trajectories
leads to the
 quantum Raychaudhuri equation (QRE) which prevents the formation of
 singularities \cite{a1}. The second order Friedmann equations obtained from the QRE
  contains a couple of quantum correction terms, the first of which can be known as cosmological constant
  while the second removes the big-bang singularity and shows that the age of our universe is infinite
  \cite{a2}. Then, it has been declared that the same result may be derived in a brane-anti-brane system;
   however, the origin of universe is N fundamental strings
   \cite{a3}. In this method, first, these strings are excited and transit to N pair
   of  D0-anti-D0-branes. These branes glue to each other and form
   a system of D5-anti-D5-branes. This system is unstable, broken
   and a pair of universe-anti-universe in additional to a wormhole is
   created. Two universes in this system interact with each other
   via the wormhole and build a BIon \cite{a4,a5,a6,a7}. Thus,
  there isn't any big bang singularity in this system and total
  age of universe is equal to sum over the age of fundamental
  string, initial system of D5-anti-D5 and present shape of
  universe. It is observed that only in the case of infinite age
  of fundamental string, the scale factor of universe becomes zero
  which means that the age of universe is infinite \cite{a3}.

  In parallel, some models in loop quantum cosmology predict that
  universe contract and expand infinite times and thus the age of
  universe may be infinite  \cite{a8}. Now, the main question
  arises that how we can unify these two types of theory in
  M-theory? We answer this question in a system of oscillating
  branes. In our model, at the beginning,  N M0-anti-M0-branes are produced from decaying of N-fundamental strings. Then,
  these branes glue to each other and  an M8-anti-M8 system is formed.  The branes in this system interact with each other, annihilate
  and two anti-M4-branes, an M4-brane,
an M3-brane plus one M0-brane are created.  The M4-brane is
compactified around a circle in eleven dimension and interacts
with other branes by exchanging M0-branes and scalars in
transverse dimensions.  The M3-brane which our universe is located
on it, wraps around the M4-brane, universe contracts and
generalized uncertainty principle on it is emerged. The M4-M3
system oscillates between anti-M4-branes and becomes close to one
of them. At this stage, the M3-brane connects to the anti-M4-brane
from one end and remains sticking and wrapping around the M4-brane
from another end. Then, the M4 rebounds, rolls, wrapped M3 opens
and universe expands. Eventually, the M4 approaches the anti-M4
and some scalars become tachyons. To solve this problem, M4 moves
away from this anti-M4-brane, rolls and M3 wraps around it again.
In these conditions, universe evolves and make a transition from
expansion phase to contraction era. These contractions and
expansions continue until infinite and thus the age of universe
may be infinite.

 The outline of the
paper is as  follows.  In section \ref{o1}, we will construct the
contraction branch of cosmology in a system of oscillating branes
and show that the origin of universe is a fundamental string . In
section \ref{o2}, we will consider expansion era of cosmology
during rolling of M4-brane in this system and estimate the age of
universe. In section \ref{o3}, we will show that by vanishing
tachyonic states, some extra energies is produced which leads to
the inflation.  In section \ref{o4}, we will indicate that the
model matches with quantum field theory prescriptions. The last
section is devoted to summary and conclusion.

\section{ First contraction branch of universe in a system of oscillating
branes }\label{o1} In this section, we will show that all
evolutions of universe from the birth to the expansion era can be
considered in a system of anti-M4-M4-M3 brane. In this model, the
formation of universe in first contraction branch is via the
process (fundamental string $\rightarrow$ M0 + anti-M0
$\rightarrow$  M8 + anti-M8 $\rightarrow$ 2M4 + compact anti-M4 +
wrapped M3 + M0 $\rightarrow$ contraction branch of universe ).
Also, in our model, some scalars make a transition to a tachyonic
phase and causes the contraction branch to be terminated.

To begin, we explain  the model of \cite{a2} in short terms. In
this mechanism, we estimate an explicit form for $\dot{H}$ = F(H),
where F(H) is a function of Hubble parameter (H) that is derived
from quantum Raychaudhuri equation . Using this function, we can
calculate the age of our world:

\begin{eqnarray}
&& \dot{H}=F(H)\rightarrow T=\frac{1}{F^{n}(H_{initial})}\int
dH\frac{1}{(H-H_{initial})^{n}} \rightarrow \infty \label{m1}
\end{eqnarray}

where $H_{initial}$ is the hubble parameter before the present era
of universe. This equation shows that the age of universe is
infinite. We will show that the same results can be obtained in
string theory. In our model, the universe is located on an
M3-brane which wraps around a compact M4 from one end and attached
to anti-M4 from another end. By oscillating and rolling M4, M3
oscillates between wrapping and opening states and consequently,
universe oscillates between contraction and expansion branches. To
show this, we  use of the mechanism in \cite{a3}. In this paper,
it has been shown that a fundamental string can decay to a pair of
D0-anti-D0-branes or a pair of M0-anti-M0-branes in additional to
some extra energy ($V$) \cite{a3}:

\begin{eqnarray}
&&  S_{F-string} = S_{D0}+S_{anti-D0}
=S_{M0}+S_{anti-M0}+2V(extra) \label{m2}
\end{eqnarray}

where, the actions of D0-brane and M0-branes have been defined as
\cite{a3,a9,a10,a11,a12,a13,a14,a15,a16,a17}:

\begin{eqnarray}
&& S_{D0} =S_{anti-D0}= -T_{D0} \int dt Tr( \Sigma_{m=0}^{9}
[X^{m},X^{n}]^{2}) \label{m3}
\end{eqnarray}

\begin{eqnarray}
S_{M0} = S_{anti-M0} = T_{M0}\int dt Tr( \Sigma_{M,N,L=0}^{10}
\langle[X^{M},X^{N},X^{L}],[X^{M},X^{N},X^{L}]\rangle) \label{m4}
\end{eqnarray}

Here, $T_{D0}$ and $T_{M0}$ are the brane tensions, $X^{m}$ are
transverse scalars,  $X^{M}=X^{M}_{\alpha}T^{\alpha}$ and

\begin{eqnarray}
 &&[T^{\alpha}, T^{\beta}, T^{\gamma}]= f^{\alpha \beta \gamma}_{\eta}T^{\eta} \nonumber \\&&\langle T^{\alpha},
  T^{\beta} \rangle = h^{\alpha\beta} \nonumber \\&& [X^{M},X^{N},X^{L}]=[X^{M}_{\alpha}T^{\alpha},X^{N}_{\beta}T^{\beta},X^{L}_{\gamma}T^{\gamma}]\nonumber \\&&\langle X^{M},X^{M}\rangle = X^{M}_{\alpha}X^{M}_{\beta}\langle T^{\alpha}, T^{\beta} \rangle
\label{m5}
\end{eqnarray}

 As can be seen from above equation, the relevant action of D0-branes contains two dimensional Nambu-Poisson bracket while the action of M3 has three one
  with the Li-3-algebra \cite{a14,a15,a16,a17}. Also, the actions of D0 and M0 have the following relations  \cite{a3}:

\begin{eqnarray}
&& S_{M0} =  S_{D0} + V_{Extra}\label{m6} \\
&& \nonumber \\
&& \text{where}\nonumber \\
&& \nonumber \\
&& V_{Extra}= -6T_{M0}\int dt
\Sigma_{M,N,L,E,F,G=0}^{9}\varepsilon_{MNLD}\varepsilon_{EFG}^{D}X^{M}X^{N}X^{L}X^{E}X^{F}X^{G}
\nonumber
\end{eqnarray}

Here
$T_{D0}=6T_{M0}(\frac{R^{2}}{l_{p}^{3}})=\frac{1}{g_{s}l_{s}}$ is
the brane tension and $g_{s}$ and $l_{s}$ are the string coupling
and string length respectively.

At this stage, we want to obtain the relevant action for Dp-brane
by summing over the actions of D0-branes. To this end, we use of
following mappings \cite{a3,a9,a10,a11,a12}:

\begin{eqnarray}
&& \Sigma_{a=0}^{p}\Sigma_{m=0}^{9}\rightarrow \frac{1}{(2\pi l_{s})^{p}}\int d^{p+1}\sigma \Sigma_{m=p+1}^{9}\Sigma_{a=0}^{p} \qquad \lambda = 2\pi l_{s}^{2} \nonumber \\
&&[X^{a},X^{i}]=i
\lambda \partial_{a}X^{i}\qquad  [X^{a},X^{b}]= i \lambda^{2} F^{ab}\nonumber \\
&& i,j=p+1,..,9\qquad a,b=0,1,...p\qquad m,n=0,1,..,9 \label{m7}
\end{eqnarray}

Now, we can obtain the relevant action of Dp-brane
\cite{a3,a9,a10,a11,a12}:

\begin{eqnarray}
&& S_{Dp} =-\Sigma_{a=0}^{p}T_{D0} \int dt Tr( \Sigma_{m=0}^{9}
[X^{m},X^{n}]^{2}) = \Sigma_{a=0}^{p}S_{D0} \nonumber \\
&&=-T_{Dp} \int d^{p+1}\sigma Tr (\Sigma_{a,b=0}^{p}
\Sigma_{i,j=p+1}^{9}
\{\partial_{a}X^{i}\partial_{b}X^{i}-\frac{1}{2
\lambda^{2}}[X^{i},X^{j}]^{2}+\frac{\lambda^{2}}{4} (F_{ab})^{2}
\}) \label{m8}
\end{eqnarray}

Also, to derive the relevant action for Mp-branes, we sum over the
action of M0-branes and use of following mappings
\cite{a3,a9,a10,a11,a12,a13,a14,a15,a16,a17}:

\begin{eqnarray}
&&\langle[X^{a},X^{b},X^{i}],[X^{a},X^{b},X^{i}]\rangle=
\frac{1}{2}\varepsilon^{abc}\varepsilon^{abd}(\partial_{a}X^{i}_{\alpha})(\partial_{a}X^{i}_{\beta})\langle(T^{\alpha},T^{\beta}\rangle
=
 \frac{1}{2}\langle \partial_{a}X^{i},\partial_{a}X^{i}\rangle \nonumber \\
&&\nonumber \\
&&\langle[X^{a},X^{b},X^{c}],[X^{a},X^{b},X^{c}]\rangle=
(F^{abc}_{\alpha\beta\gamma})(F^{abc}_{\alpha\beta\eta})\langle[T^{\alpha},T^{\beta},T^{\gamma}],[T^{\alpha},T^{\beta},T^{\eta}]\rangle)=\nonumber \\
&&
(F^{abc}_{\alpha\beta\gamma})(F^{abc}_{\alpha\beta\eta})f^{\alpha
\beta \gamma}_{\sigma}h^{\sigma \kappa}f^{\alpha \beta
\eta}_{\kappa} \langle T^{\gamma},T^{\eta}\rangle=
(F^{abc}_{\alpha\beta\gamma})(F^{abc}_{\alpha\beta\eta})\delta^{\kappa
\sigma} \langle T^{\gamma},T^{\eta}\rangle=
\langle F^{abc},F^{abc}\rangle\nonumber \\
&&\nonumber \\
&&\Sigma_{m}\rightarrow \frac{1}{(2\pi)^{p}}\int d^{p+1}\sigma
\Sigma_{m-p-1} i,j=p+1,..,10\quad a,b=0,1,...p\quad m,n=0,..,10~~ \nonumber \\
&& F_{abc}=\partial_{a} A_{bc}-\partial_{b} A_{ca}+\partial_{c}
A_{ab} \label{m9}
\end{eqnarray}

Using above relations, the action of  Mp-brane  can be obtained as
 \cite{a3,a14,a15,a16,a17}:

\begin{eqnarray}
&& S_{Mp} = \Sigma_{a=0}^{p}S_{M0}=-\Sigma_{a=0}^{p}T_{M0} \int dt
Tr( \Sigma_{m=0}^{9}
\langle[X^{a},X^{b},X^{c}],[X^{a},X^{b},X^{c}]\rangle) =
\nonumber \\ && -T_{Mp} \int d^{p+1}\sigma Tr
(\Sigma_{a,b,c=0}^{p} \Sigma_{i,j,k=p+1}^{10}
\{\langle\partial_{a}X^{i},\partial_{a}X^{i}\rangle
-\frac{1}{4}\langle[X^{i},X^{j},X^{k}],[X^{i},X^{j},X^{k}]\rangle+\nonumber
\\ &&\frac{1}{6} \langle F_{abc},F_{abc}\rangle \}) \label{m10}
\end{eqnarray}

 Now, we can build our model in M-theory. For this, first, a pair of M8-anti-M8-branes are constructed from joining M0-branes. The, these objects decay to
 two anti-M4-branes, one M4-brane, one M3-brane in additional one M0-brane:

\begin{eqnarray}
&& S_{tot}= \Sigma_{a=0}^{8}S_{M0} + \Sigma_{a=0}^{8}S_{anti-M0}=
S_{M8} + S_{anti-M8}=\nonumber
\\&&2\Sigma_{a=0}^{4}S_{anti-M0} + \Sigma_{a=0}^{4}S_{M0} + \Sigma_{a=0}^{3}S_{M0} + S_{M0} =\nonumber
\\&& 2S_{anti-M4} +
S_{M4} + S_{M3} + S_{M0} \label{m11}
\end{eqnarray}

In our method, the M4-brane is compactified around a circle in
eleven dimensions and M3 wraps around it. Then, the M4-M3 system
moves toward one of anti-M4-branes. Approaching anti-M4, M3 sticks
to the anti-M4 from one end and remains sticking to M4 from
another end. By getting away the M4, it rolls and M3 opens. This
process repeat many times. During wrapping and compactifications,
some components of gauge field should be replaced by scalars and
by opening, some scalars converts to gauge field
\cite{a3,a9,a10,a11,a12,a13,a14,a15,a16,a17}. For this reason, we
write following relations for mappings which includes both of
states :

\begin{eqnarray}
&& [X^{a},X^{b},X^{c/i}]=\alpha
\partial_{a}\partial_{b}X^{i} + \beta F_{abc}  \nonumber
\\&&
[X^{a},X^{i/b},X^{j/c}]=\alpha
(X^{i}\partial_{a}X^{j}+X^{j}\partial_{a}X^{i})+\beta F_{abc}
\nonumber
\\&& [X^{i},X^{j},X^{k}]=
\varepsilon^{\alpha\beta\gamma}X^{i}_{\beta}X^{j}_{\beta}X^{k}_{\gamma}
\label{m12}
\end{eqnarray}

where a,b,c are the indexes on the branes and i,j,k are indexes in
transverse directions. By wrapping and compactification of branes,
some of brane's indexes like a,b or c are replaced by i,j. Also,
$\alpha$ and $\beta$ are functions of time which increase and
decrease during wrapping and opening phases. When, M3 wraps around
M4 completely, $\alpha$ should be maximum and as M3 opens
completely, $\beta$ is maximum. For this reason, we suggest that
$\alpha=sin\omega t$ and $\beta=cos\omega t$ where $\omega$ is the
frequency of oscillation ($\omega=\frac{2\pi}{T}$) and T is the
time of period. Using equations (\ref{m11}) and (\ref{m12}), we
obtain:

\begin{eqnarray}
&& S_{M3-M4} = -T_{D3} \int d^{4}\sigma Tr (\Sigma_{a,b=0}^{3}
\Sigma_{i,j=4}^{9} \{\partial_{a}X^{i}\partial_{b}X^{j}+
\alpha^{2}
\partial_{a}\partial_{b}X^{i}\partial_{a}\partial_{b}X^{i} + \beta^{2} F_{abc}^{2}+\beta^{3} F_{abc}^{3}\nonumber \\&&
+\alpha \beta \partial_{a}\partial_{b}X^{i} F_{abc}+ \alpha^{2}
(X^{i}X^{j}\partial_{a}X^{j}\partial_{b}X^{i})+\alpha F_{abc}
(X^{i}\partial_{a}X^{j}+X^{j}\partial_{a}X^{i})+\alpha^{2}\beta
\partial_{a}\partial_{b}X^{i}\partial_{a}\partial_{b}X^{i}F_{abc}\nonumber
\\&&
 +\varepsilon^{\alpha\beta\gamma}\varepsilon^{\alpha'\beta'\gamma'}X^{i}_{\alpha}X^{j}_{\beta}X^{k}_{\gamma}X^{i}_{\alpha'}X^{j}_{\beta'}X^{k}_{\gamma'} \})
 =S_{M3}+V_{int}
  \label{m13}
\end{eqnarray}

where $V_{int}=-T_{D3} \int d^{4}\sigma Tr (\Sigma_{a,b=0}^{3}
\Sigma_{i,j=4}^{9} \{\beta^{3} F_{abc}^{3}+\alpha^{2}\beta
\partial_{a}\partial_{b}X^{i}\partial_{a}\partial_{b}X^{i}F_{abc}\}$ is the interaction potential between M3 and M4. Using the above equation, we can derive the
equation of motion for $X^{i}$:

\begin{eqnarray}
&& \{\partial_{a}^{2}+(\alpha^{2}+2\alpha\beta
A_{a})\partial_{a}^{4}+\alpha^{2}\beta
A_{a}X^{j}\partial_{a}^{6}+\alpha^{2}X^{j}X^{k}X^{l}\partial_{a}^{2}+\alpha
A_{a}\partial_{a}^{3} + \frac{\partial^{2} V}{\partial
x^{2}}\}X^{i}=0 \nonumber \\&&
V=\varepsilon^{\alpha\beta\gamma}\varepsilon^{\alpha'\beta'\gamma'}X^{i}_{\alpha}X^{j}_{\beta}X^{k}_{\gamma}X^{i}_{\alpha'}X^{j}_{\beta'}X^{k}_{\gamma'}
 \label{m14}
\end{eqnarray}

At this stage, we substitute $\partial_{a} = p_{a}$ and
$\frac{\partial^{2} V}{\partial x^{2}} = m^{2}$ in above equation
and obtain:

\begin{eqnarray}
&&\{p_{a}^{2}+(\alpha^{2}+2\alpha\beta
A_{a})p_{a}^{4}+\alpha^{2}\beta
A_{a}X^{j}p_{a}^{6}+\alpha^{2}X^{j}X^{k}X^{l}p_{a}^{2}-\alpha
A_{a}p_{a}^{3} + m^{2}\}X^{i}=0 \label{m15}
\end{eqnarray}

If we compare above equations with usual equation for scalar
fields,

\begin{eqnarray}
&& \{\bar{p}_{\alpha}^{2}+ m^{2}\}X^{i}=0
 \label{m16}
\end{eqnarray}

we can define the momentum $\bar{p}$  as:

\begin{eqnarray}
&& \bar{p}_{a}\sim(1+\alpha^{2}X^{j}X^{k}X^{l})^{1/2}p_{a}
-(\alpha^{2}+2\alpha\beta A_{a})^{1/2}p_{a}^{2}+\alpha^{2}\beta
A_{a}X^{j}p_{a}^{3}
 \label{m17}
\end{eqnarray}

Thus our model produces the commutation relations in GUP:

\begin{eqnarray}
&& \{{x}_{a},\bar{p}_{b}\}=
(1+\alpha^{2}X^{j}X^{k}X^{l})^{1/2}\delta_{ab} -
(\alpha^{2}+4\alpha\beta A_{a})^{1/2}p_{a}+3\alpha^{2}\beta
A_{a}X^{j}p_{a}^{2} \label{m18}
\end{eqnarray}

This is the GUP proposed in \cite{a18,a19,a20,a21,a22} which
predicts maximum observable momenta besides the existence of
minimal measurable length and is consistent with Doubly Special
Relativity (DSR) theories, String Theory and Black Holes Physics.
Thus, wrapping M3 around M4 leads to the non-commutative relations
in GUP. Equation(\ref{m18})  gives the following uncertainty with
using the argument used in \cite{a23,a24,a25}:

\begin{eqnarray}
&& \Delta x \Delta p \geq
\frac{1}{2}[(1+\alpha^{2}X^{j}X^{k}X^{l})^{1/2} -
(\alpha^{2}+4\alpha\beta A_{a})^{1/2}p_{a}+3\alpha^{2}\beta
A_{a}X^{j}p_{a}^{2}]\label{m19}
\end{eqnarray}

The solution of  the above inequality as quadratic equation in
$\Delta p$ is \cite{a23,a24,a25}:

\begin{eqnarray}
&& \Delta p \geq \frac{2\Delta x+(\alpha^{2}+4\alpha\beta
A_{a})^{1/2}}{3\alpha^{2}\beta
A_{a}X^{j}}[1-\sqrt{1-\frac{6\alpha^{2}\beta A_{a}X^{j}}{(2\Delta
x+(\alpha^{2}+4\alpha\beta A_{a})^{1/2})^{2}}}]\label{m20}
\end{eqnarray}

Now, we assume that wrapped M3-brane could be modelled as (D -1)-
dimensional sphere of size equal to twice of Schwarzschild radius,
$r_{s}$. Thus the uncertainty in position of particle has its
minimum value given by \cite{a23,a24,a25}:

\begin{eqnarray}
&& \Delta x = 2r_{s}=
2\lambda_{D}[\frac{G_{D}m}{c^{2}}]^{\frac{1}{D-3}}\label{m21}
\end{eqnarray}

where
$\lambda_{D}=[\frac{16\pi}{(D-2)\Omega_{D-2}}]^{\frac{1}{D-3}}$
and
$\Omega_{D}=\frac{2\pi^{\frac{D-1}{2}}}{\Gamma(\frac{D-1}{2})}$.
We substitute the position defined by (\ref{m21}) in equation
(\ref{m20})and derive $\Delta p$ as :

\begin{eqnarray}
&& \Delta p \geq
\frac{4\lambda_{D}[\frac{G_{D}m}{c^{2}}]^{\frac{1}{D-3}}+(4\alpha\beta
A_{a})^{1/2}}{3\alpha^{2}\beta
A_{a}X^{j}}[1-\sqrt{1-\frac{6\alpha^{2}\beta
A_{a}X^{j}}{(4\lambda_{D}[\frac{G_{D}m}{c^{2}}]^{\frac{1}{D-3}}+(\alpha^{2}+4\alpha\beta
A_{a})^{1/2})^{2}}}]\label{m22}
\end{eqnarray}

Applying definition of mass in equation (\ref{m14}), we calculate
the explicit form of mass:

\begin{eqnarray}
&& m^{2} = \frac{\partial^{2} V}{\partial
x^{2}}=\varepsilon^{\alpha\beta\gamma}\varepsilon^{\alpha'\beta'\gamma'}\frac{\partial^{2}
(X^{i}_{\alpha}X^{j}_{\beta}X^{k}_{\gamma}X^{i}_{\alpha'}X^{j}_{\beta'}X^{k}_{\gamma'})}{\partial
x^{2}} \label{m23}
\end{eqnarray}

Substituting equation (\ref{m23}) in equation (\ref{m22}), we get:

\begin{eqnarray}
&& \Delta p \geq
\frac{4\lambda_{D}[\frac{G_{D}\varepsilon^{\alpha\beta\gamma}\varepsilon^{\alpha'\beta'\gamma'}\frac{\partial^{2}
(X^{i}_{\alpha}X^{j}_{\beta}X^{k}_{\gamma}X^{i}_{\alpha'}X^{j}_{\beta'}X^{k}_{\gamma'})}{\partial
x^{2}} }{c^{2}}]^{\frac{1}{2D-6}}+(4\alpha\beta
A_{a})^{1/2}}{3\alpha^{2}\beta A_{a}X^{j}}\times
\nonumber\\&&[1-\sqrt{1-\frac{6\alpha^{2}\beta
A_{a}X^{j}}{(4\lambda_{D}[\frac{G_{D}\varepsilon^{\alpha\beta\gamma}\varepsilon^{\alpha'\beta'\gamma'}\frac{\partial^{2}
(X^{i}_{\alpha}X^{j}_{\beta}X^{k}_{\gamma}X^{i}_{\alpha'}X^{j}_{\beta'}X^{k}_{\gamma'})}{\partial
x^{2}} }{c^{2}}]^{\frac{1}{2D-6}}+(\alpha^{2}+4\alpha\beta
A_{a})^{1/2})^{2}}}]\label{m24}
\end{eqnarray}

This equation yields the following inequality for the scalars in
transverse direction:

\begin{eqnarray}
&& A_{a}  \leq [\frac{
(4\lambda_{D}[\frac{G_{D}\varepsilon^{\alpha\beta\gamma}\varepsilon^{\alpha'\beta'\gamma'}\frac{\partial^{2}
(X^{i}_{\alpha}X^{j}_{\beta}X^{k}_{\gamma}X^{i}_{\alpha'}X^{j}_{\beta'}X^{k}_{\gamma'})}{\partial
x^{2}} }{c^{2}}]^{\frac{1}{2D-6}}+(\alpha^{2}+4\alpha\beta
A_{a})^{1/2})^{2}}{6\alpha^{2}\beta X^{i} }] \label{m25}
\end{eqnarray}

As can be seen from the above inequality, as the M3 brane
approaches the anti-M4-brane at $t=\frac{T}{4}$, scalars on the M3
grow, the right hand of this equality becomes smaller than left
hand and inequality violates. To avoid this violation and negative
values under the $\sqrt{}$ , the square mass of some scalars
becomes negative ($m^{2}\rightarrow -m^{2}$), they transit to
tachyonic phases and contraction branch ends.

\section{Estimating the age of universe in a system of oscillating branes  }\label{o2}
In previous section, we observed that as the M3-M4 becomes close
to anti-M4, M3 attaches to it from one end and stays sticking from
another end and the system faces some tachyonic states. To remove
these states, M4 rebounds, rolls, M3 opens and expansion phase
begins. During this new phase, gauge fields ($A_{a}$) on the brane
grow, the left hand of equation (\ref{m25})becomes bigger than
right hand and inequality violates again. To avoid the negative
values under the the $\sqrt{}$ , the square mass of some scalars
becomes negative ($m^{2}\rightarrow -m^{2}$) and they become
tachyons. To solve this problem, M4 rebounds again, M3 wraps
around it and contraction epoch begins. Now, the question arises
that what is the age of universe? To answer this question, we
should calculate the contribution of branes on four dimensional
universe and write energy-momentum tensors. Using the action in
(\ref{m13}), we can calculate the energy-momentum of M3  and put
it equal to energy-momentum of universe:

\begin{eqnarray}
&& \rho = \frac{3H^{2}}{\kappa^{2}} =
\frac{1}{2}(1+\alpha^{2}(X^{j})^{2}+2\alpha
F_{abc}X^{j})(\dot{X}^{i})^{2}+\alpha^{2}
\partial_{a}\partial_{b}X^{i}\partial_{a}\partial_{b}X^{i} \nonumber \\&&
+\alpha \beta \partial_{a}\partial_{b}X^{i}
F_{abc}+\alpha^{2}\beta
\partial_{a}\partial_{b}X^{i}\partial_{a}\partial_{b}X^{i}F_{abc}+ \beta^{2} F_{abc}^{2}+\beta^{3} F_{abc}^{3}\nonumber
\\&&
 +\varepsilon^{\alpha\beta\gamma}\varepsilon^{\alpha'\beta'\gamma'}X^{i}_{\alpha}X^{j}_{\beta}X^{k}_{\gamma}X^{i}_{\alpha'}X^{j}_{\beta'}X^{k}_{\gamma'}
\nonumber\\&& p=-\frac{1}{\kappa^{2}}(3H^{2}+2\dot{H})=
 \frac{1}{2}(1+\alpha^{2}(X^{j})^{2}+2\alpha
F_{abc}X^{j})(\dot{X}^{i})^{2}\nonumber\\&&-\alpha^{2}
\partial_{a}\partial_{b}X^{i}\partial_{a}\partial_{b}X^{i}
-\alpha \beta \partial_{a}\partial_{b}X^{i}
F_{abc}-\alpha^{2}\beta
\partial_{a}\partial_{b}X^{i}\partial_{a}\partial_{b}X^{i}F_{abc}\nonumber
\\&&- \beta^{2} F_{abc}^{2}-\beta^{3} F_{abc}^{3}
 -\varepsilon^{\alpha\beta\gamma}\varepsilon^{\alpha'\beta'\gamma'}X^{i}_{\alpha}X^{j}_{\beta}X^{k}_{\gamma}X^{i}_{\alpha'}X^{j}_{\beta'}X^{k}_{\gamma'}
\label{m26}
\end{eqnarray}

Solving equations (\ref{m15},\ref{m16},\ref{m25} and
\ref{m26})simultaneously, we obtain the explicit form of
$X^{i}$,$A^{i}$ and $a(t)$:

\begin{eqnarray}
&& X^{i}\sim sin(\bar{\omega}t) \quad
\bar{\omega}=\sqrt{\omega^{2}+(\alpha^{2}+2\alpha)\omega^{4}+\alpha^{2}\beta
\omega^{6}+\alpha^{5}\omega^{2}+\omega^{3}+m^{2}} \nonumber\\&&
\nonumber\\&& A^{a}\sim
[\frac{\sqrt{G_{D}[\omega^{2}+(\alpha^{2}+2\alpha)\omega^{4}+\alpha^{2}\beta
\omega^{6}+\alpha^{5}\omega^{2}+\omega^{3}+m^{2}][30sin^{4}(\bar{\omega}t)cos^{2}(\bar{\omega}t)-6sin^{6}(\bar{\omega}t)]}+sin\omega
t}{6\alpha^{2}\beta sin(\bar{\omega}t) }]\nonumber\\&&
\nonumber\\&& a(t)\sim e^{\omega t +\int dt G(t)}\nonumber\\&&
\nonumber\\&& G(t)\sim
([\frac{\sqrt{G_{D}[\omega^{2}+(\alpha^{2}+2\alpha)\omega^{4}+\alpha^{2}\beta
\omega^{6}+\alpha^{5}\omega^{2}+\omega^{3}+m^{2}]}
\sqrt{\omega^{2}+(\alpha^{2}+2\alpha)\omega^{4}+\alpha^{2}\beta
\omega^{6}+\alpha^{5}\omega^{2}+\omega^{3}+m^{2}}}{3\alpha \beta
\sqrt{[30sin^{4}(\bar{\omega}t)cos^{2}(\bar{\omega}t)-6sin^{6}(\bar{\omega}t)]}
}])\times \nonumber\\&&
[120sin^{3}(\bar{\omega}t)cos^{3}(\bar{\omega}t)+30sin^{5}(\bar{\omega}t)cos(\bar{\omega}t)-30cos(\bar{\omega}t)sin^{5}(\bar{\omega}t)]+\alpha^{2}
sin^{2}(\bar{\omega}t))cos^{2}(\bar{\omega}t) +
cos(2\bar{\omega}t)+\nonumber\\&&[\frac{cos\omega
t\sqrt{G_{D}[\omega^{2}+(\alpha^{2}+2\alpha)\omega^{4}+\alpha^{2}\beta
\omega^{6}+\alpha^{5}\omega^{2}+\omega^{3}+m^{2}][30sin^{4}(\bar{\omega}t)cos^{2}(\bar{\omega}t)-6sin^{6}(\bar{\omega}t)]}+2cos\omega
t sin\omega t}{3\alpha^{2}\beta sin^{3}(\bar{\omega}t) }]
\label{m27}
\end{eqnarray}

As can be seen from these equations, during contraction
branch($0<t<\frac{T}{4}$), the scalar fields ($X^{i}$) grow; while
the gauge fields ($A^{a}$) decrease. However, by passing time and
opening M3, universe enters into expansion phase
($\frac{T}{4}<t<\frac{T}{2}$), gauge fields grow and scalars
decreases. Also, this equation shows that only in the case that
time be infinite ($t \rightarrow -\infty $), the scale factor
becomes zero. This means that the age of universe is infinite and
thus our result is consistent with results of \cite{a2}.

\section{ Considering the inflation era at the beginning of expansion branch }\label{o3}
Until now, we have shown that by wrapping and opening M3, universe
contracts and expands. Also, we have replace the big bang
singularity by a fundamental string and indicate that the age of
universe is infinite. Now, another question arises that how our
universe undergoes an inflation phase at the beginning of
expansion branch? To reply this question, we remind that at the
end of contraction, some scalars gain negative square mass and
transit to tachyons. To remove these states, contraction stops and
universe enters to an expansion epoch. Also, the negative square
mass of scalars ($-m^{2}=-\frac{\partial^{2}V}{\partial x^{2}}$)
at the end of contraction should be converted to the positive
square mass ($m^{2}=\frac{\partial^{2}V}{\partial x^{2}}$))at the
beginning of expansion. Thus, the energy of system changes and
some energy
 is released. Using equation (\ref{m14}), we can get:

\begin{eqnarray}
&&m^{2}-(-m^{2})=2m^{2} \quad V=V_{\text{end of
contraction}}=V_{\text{beginning of expansion}} \rightarrow
\nonumber\\&& \frac{\partial^{2}V_{\text{beginning of
expansion}}}{\partial x^{2}}-(-\frac{\partial^{2}V_{\text{end of
contraction}}}{\partial x^{2}})=2\frac{\partial^{2}V}{\partial
x^{2}}\rightarrow  \nonumber\\&& V_{inf}=V_{\text{beginning of
expansion}}-V_{\text{end of contraction}}=2V=2
\varepsilon^{\alpha\beta\gamma}\varepsilon^{\alpha'\beta'\gamma'}X^{i}_{\alpha}X^{j}_{\beta}X^{k}_{\gamma}X^{i}_{\alpha'}X^{j}_{\beta'}X^{k}_{\gamma'}
 \label{m28}
\end{eqnarray}

This energy causes that the velocity of opening of M3 increases
and our universe which is located on this brane, experience an
inflation phase.
  Using equations (\ref{m27}and\ref{m28}) and assuming that inflation starts at the beginning of expansion ($t=\frac{T}{4}$), we can obtain the
Hubble parameter:

\begin{eqnarray}
&&\rho_{total}=\rho+\rho_{inf}\nonumber\\&&\rho_{inf} =
\frac{3H_{inf}^{2}}{\kappa^{2}}=V_{inf}
=2V=2sin^{6}(\bar{\omega}t)\rightarrow \nonumber\\&&
H_{inf}=\sqrt{\frac{2}{3}}\kappa
sin^{3}(\bar{\omega}t)\nonumber\\&&
t=\frac{T}{4}+t_{inf}\rightarrow H_{inf}=\sqrt{\frac{2}{3}}\kappa
cos^{3}(\bar{\omega}t_{inf}) \nonumber\\&&
H_{tot}^{2}=H^{2}+H_{inf}^{2} \quad H=G+\omega \nonumber\\&&
H_{tot}^{2}=(G+\omega)^{2}+\frac{2\kappa}{3}
cos^{6}(\bar{\omega}t_{inf}) \label{m29}
\end{eqnarray}

We can test our model by calculating the magnitude of the
slow-roll parameters and the tensor-to- scalar ratio r defined in
\cite{a26} and comparing with previous predictions:

\begin{eqnarray}
&&\varepsilon=-\frac{\dot{H}_{tot}}{H_{tot}^{2}}=\frac{2\dot{G}(G+\omega)+\frac{12\kappa}{3}
\bar{\omega}cos^{5}(\bar{\omega}t_{inf})sin(\bar{\omega}t_{inf})}{[(G+\omega)^{2}+\frac{2\kappa}{3}
cos^{6}(\bar{\omega}t_{inf})]^{3/2}}\nonumber\\&& \eta
=-\frac{\ddot{H}_{tot}}{2H_{tot}\dot{H }_{tot}}=[\frac{2\ddot{G}(G+\omega)+2\dot{G}\dot{G}+\frac{12\kappa}{3}
\bar{\omega}cos^{6}(\bar{\omega}t_{inf})-\frac{60\kappa}{3}
\bar{\omega}cos^{4}(\bar{\omega}t_{inf})sin^{2}(\bar{\omega}t_{inf})}{[(G+\omega)^{2}+\frac{2\kappa}{3}
cos^{6}(\bar{\omega}t_{inf})]^{1/2}}\nonumber\\&&+\frac{(2\dot{G}(G+\omega)+\frac{12\kappa}{3}\bar{\omega}
cos^{5}(\bar{\omega}t_{inf})sin(\bar{\omega}t_{inf}))^{2}}{[(G+\omega)^{2}+\frac{2\kappa}{3}
cos^{6}(\bar{\omega}t_{inf})]^{3/2}}]\times
\frac{1}{2\dot{G}(G+\omega)+\frac{12\kappa}{3}\bar{\omega}
cos^{5}(\bar{\omega}t_{inf})sin(\bar{\omega}t_{inf})}\nonumber\\&&
\nonumber\\&&t_{inf}\ll T \quad and \quad \bar{\omega}\sim
\frac{1}{T^{3}}\rightarrow \bar{\omega}t_{inf}\ll1\rightarrow
\nonumber\\&&
 sin(\bar{\omega}t_{inf})\ll cos(\bar{\omega}t_{inf})\quad and
 \quad
sin(\bar{\omega}t_{inf})\sim \frac{t_{inf}}{T}\sim 0 \quad and
\quad cos(\bar{\omega}t_{inf})\sim 1\Rightarrow \nonumber\\&&
\varepsilon=\frac{1}{G^{2}} \ll 1 \quad \eta \sim \frac{1}{G} \ll
1 \Rightarrow \quad r=16 \varepsilon\sim\frac{16}{G^{2}} \ll 1
 \label{m30}
\end{eqnarray}

This equation shows that slow parameters are very small and thus
our model confirms the prediction of previous models for inflation
era in \cite{a26}. Another interesting result that comes out from
this equation is the value of the tensor-to- scalar ratio r which
is very smaller than one and is in agreement with  experimental
data \cite{a27}. Thus, the extra energy which is produced during
vanishing tachyons, leads to an increase in velocity of expansion
and occurring inflation.

\section{ Reducing the model to quantum field theory prescriptions in four dimensional universe }\label{o4}
In this section, we will show that by reducing field theory in
eleven dimensional M-theory to field theory in  four dimensional
universe, our model matches with known models in gravity. To this
end, we use of following relations:

\begin{eqnarray}
&& \int d^{4}\sigma  F_{abc}^{2}=\int d^{4}\sigma (\partial_{a}
A_{bc}-\partial_{b} A_{ca}+\partial_{c}
A_{ab})(F_{abc})=\nonumber\\&&-\int d^{4}\sigma
(A_{bc}\partial_{a}F_{abc} -A_{ca}\partial_{b}F_{abc} +
A_{ab}\partial_{c}F_{abc}) +\int d^{4}\sigma
\partial_{a}(O)\nonumber\\&& \int d^{4}\sigma  F_{abc}^{3}=\int d^{4}\sigma (\partial_{a}
A_{bc}-\partial_{b} A_{ca}+\partial_{c}
A_{ab})(F_{abc}^{2})=\nonumber\\&&-\int d^{4}\sigma
(A_{bc}\partial_{a}F_{abc} -A_{ca}\partial_{b}F_{abc} +
A_{ab}\partial_{c}F_{abc})F_{abc} +\int
d^{4}\partial_{a}(O)\nonumber\\&& \int d^{4}\sigma
A_{bc}F_{abc}\partial_{a}F_{abc}=\int d^{4}\sigma
(\partial_{b}A_{c}-\partial_{c}A_{b})F_{abc}\partial_{a}F_{abc}=-\int
d^{4}\sigma
(A_{c}\partial_{b}F_{abc}-A_{b}\partial_{c}F_{abc})\partial_{a}F_{abc}\nonumber\\&&
\int d^{4}\sigma
\partial_{\alpha}\partial_{\beta}X^{i}F_{abc}=-\int d^{4}\sigma
\partial_{\beta}X^{i}\partial_{\alpha}F_{abc}+\int d^{4}\sigma
\partial_{\alpha}O
\label{m31}
\end{eqnarray}

Substituting above relations in action (\ref{m13}), we obtain:

\begin{eqnarray}
&& S_{M3-M4} = -T_{D3} \int d^{4}\sigma Tr (\Sigma_{a,b=0}^{3}
\Sigma_{i,j=4}^{9} \{\partial_{a}X^{i}\partial_{b}X^{j}+
\alpha^{2}
\partial_{a}\partial_{b}X^{i}\partial_{a}\partial_{b}X^{i} - \beta^{2} A_{ab}\partial_{a}F_{abc}+\beta^{3} A_{a}\partial_{b}F_{abc}\partial_{c}F_{abc}\nonumber \\&&
-\alpha \beta \partial_{a}X^{i} \partial_{b}F_{abc}+ \alpha^{2}
(X^{i}X^{j}\partial_{a}X^{j}\partial_{b}X^{i})+\alpha
(X^{i}X^{j})\partial_{a}F_{abc} +\alpha^{2}\beta
\partial_{a}\partial_{b}X^{i}\partial_{b}X^{i}\partial_{a}F_{abc}\nonumber
\\&&
 +\varepsilon^{\alpha\beta\gamma}\varepsilon^{\alpha'\beta'\gamma'}X^{i}_{\alpha}X^{j}_{\beta}X^{k}_{\gamma}X^{i}_{\alpha'}X^{j}_{\beta'}X^{k}_{\gamma'}
 -\beta^{3} A_{a}F_{abc}\partial_{c}\partial_{b}F_{abc} -\alpha^{2}\beta
\partial_{a}X^{i}\partial_{b}X^{i}\partial_{a}\partial_{b}F_{abc}\})
  \label{m32}
\end{eqnarray}

At this stage, we can show that this action matches to action in
four dimensional field theory by using following mappings:

\begin{eqnarray}
&& X^{i}\rightarrow \phi \quad A_{ab}\rightarrow h_{ab} \quad
A_{a} \rightarrow e_{a}\nonumber\\&&  h_{ab}=\sqrt{-g}h^{ab} \quad
g_{ab}=\eta_{ab}+h_{ab}\nonumber\\&& F_{abc}=\partial_{a}
A_{bc}-\partial_{b} A_{ca}+\partial_{c} A_{ab}\rightarrow
\nonumber\\&& F_{abc}=\partial_{a} h_{bc}-\partial_{b}
h_{ca}+\partial_{c} h_{ab}\rightarrow
\nonumber\\&&\partial_{a}F_{abc}=\partial_{a}^{2}
h_{bc}-\partial_{a}\partial_{b} h_{ca}+\partial_{a}\partial_{c}
h_{ab}\rightarrow \nonumber\\&& g^{bc}F_{abc}=\partial_{a}^{2}
h_{b}^{c}+..\rightarrow \sqrt{-g}R
  \label{m33}
\end{eqnarray}

where $\phi$ is the scalar field, $h_{ab}$ is the graviton field
and $g_{ab}$ is the component of metric. Replacing strings and
three form fields by scalars and elements of metric in four
dimensions, we get:

\begin{eqnarray}
&& S_{\text{field theory}} = -T_{D3} \int d^{4}\sigma Tr
(\Sigma_{a,b=0}^{3} \Sigma_{i,j=4}^{9}
\{\partial_{a}\phi\partial_{b}\phi+ \alpha^{2}
\partial_{a}\partial_{b}\phi\partial_{a}\partial_{b}\phi - \beta^{2}\sqrt{-g}R+\beta^{3} e_{a}\sqrt{-g}R^{2}\nonumber \\&&
-\alpha \beta \sqrt{-g} \partial_{a}\phi R+ \alpha^{2}
\phi^{2}\partial_{a}\phi\partial_{b}\phi+\alpha \phi^{2}
\sqrt{-g}R +\alpha^{2}\beta
\partial_{a}\partial_{b}\phi\partial_{b}\phi \sqrt{-g}R
 -\beta^{3} e_{a}\sqrt{g}\partial_{a}h_{ab}\partial_{b}R+\nonumber \\&&\phi^{6}
  -\alpha^{2}\beta \sqrt{-g}
\partial_{a}\phi\partial_{b}\phi\partial_{a}R\})
  \label{m34}
\end{eqnarray}

Now, we can rewrite the above action as follows:

\begin{eqnarray}
&& S_{\text{field theory}} = -T_{D3} \int d^{4}\sigma
\{\sqrt{-g}F(R,\phi)+\partial_{a}\phi\partial_{b}\phi+V(\phi)\}
  \label{m34}
\end{eqnarray}

where

\begin{eqnarray}
&& F(R,\phi)=( - \beta^{2}-\alpha
\beta\partial_{a}\phi+\alpha^{2}\beta
\partial_{a}\partial_{b}\phi\partial_{b}\phi)R+\beta^{3}
e_{a}R^{2}+(-\alpha^{2}\beta
\partial_{a}\phi\partial_{b}\phi -\beta^{3}
e_{a}\partial_{a}h_{ab})\partial_{a}R\nonumber\\&&
V(\phi)=\phi^{6}+ \alpha^{2}
\phi^{2}\partial_{a}\phi\partial_{b}\phi+ \alpha^{2}
\partial_{a}\partial_{b}\phi\partial_{a}\partial_{b}\phi
  \label{m35}
\end{eqnarray}

This equation is very the same of actions in $F(R)$ gravity which
has been discussed in \cite{a28}. This means that by redefining
the quantum fields in M-theory and obtaining their relations by
fields in four dimensional universe, the action of model matches
the relevant action in quantum field theory prescription.

\section{Summary and Discussion} \label{sum}
In this research, we have reconsidered the results of \cite{a2} in
a system of oscillating brane. We have discussed that the universe
contracts and expands as due to interaction between branes. In our
model, first, N fundamental string transit to N pairs of
M0-anti-M0-brane. Then, these branes glue to each other and build
a pair of M8-anti-M8-branes. This system is unstable, broken and
two anti-M4-branes, an M4-brane, an M3-brane and an M0-brane are
produced. M4-branes is compactified around a circle and M3 which
our universe is located on it; wraps around it. The system of
M4-M3 is located between anti-M4-branes and oscillate. As this
system becomes close to one of anti-M4 branes, the M3 attaches to
it from one end and stay sticking to another from another end.
Also, the square mass of some scalars becomes negative and they
make a transition to tachyonic states. To remove these states, M4
rebounds, rolls, M3 opens and expansion branch of universe begins.
When M4 approaches to another anti-M4-brane, some other scalars
gain negative square mass and new phase of tachyon is created. To
solve this problem, M4 rebounds again, M3 wraps around it and new
contraction branch starts. We compare thee energy-momentum tensor
derived in this model with the energy-momentum tensor in our
present stage of universe and obtain the scale factor. We notice
that this scale factor, only in the case of $t\rightarrow -\infty$
becomes zero. This means that the age of universe may be infinite
which is consistent with prediction of \cite{a2}. Also, we show
that by disappearing tachyonic states, some energy is produced
which leads to an acceleration in openning of M3 and expansion of
universe. Finally, by reducing the quantum fields in eleven
dimensional M-theory to ones in four dimensional universe, we
observe that our model is consistent with usual field theory.

\section*{Acknowledgments}
\noindent  A. Sepehri would like to thank of the Research
Institute for Astronomy and Astrophysics of Maragha, Iran for
financial support during investigation in this work.  We are very
grateful of Ali Mohammad for his lecture in cosmology that gives
us new insight into this subject. We also thank of referee for
nice comments that help us to improve our paper.

 \end{document}